\documentclass{ws-ijmpd}

\begin{document}

\markboth{Authors' Names}
{Instructions for Typing Manuscripts (Paper's Title)}

%
\catchline{}{}{}{}{}
%

\title{BOSE-EINSTEIN CONDENSATION OF CHARGED PARTICLES AND SELF-MAGNETIZATION}

\author{H. P\'{E}REZ ROJAS}

\address{ICIMAF, Calle E No. 309, La Habana,10400, Cuba\\
hugo@icmf.inf.cu}

\author{A. P\'{E}REZ MART\'{I}NEZ}

\address{ICIMAF, Calle E No. 309, La Habana,10400, Cuba\\
aurora@icmf.inf.cu}

\author{H. MOSQUERA CUESTA}
\address{Centro Brasileiro de
Pesquisas F\'{\i}sicas, Laborat\'orio de Cosmologia e F\'{\i}sica
Experimental de Altas Energias Rua\\
Dr. Xavier Sigaud 150, Cep 22290-180, Urca, Rio de Janeiro, RJ,
Brazil \\ hermanjc@cbpf.br} 

\maketitle

\begin{history}
\received{(Day Month Year)}
\revised{(Day Month Year)}
\end{history}

\begin{abstract}
We discuss the Bose-Einstein condensation of relativistic vector
charged particles in a strong external magnetic field in very
dense matter, as may be paired spin-up electrons. We show that for
electrons such systems may maintain self-consistently magnetic
field strengths in the interval $10^{10}-10^{13}$ Gauss.
This could be the origin of large magnetic fields in some white
dwarfs, but may also impose bounds due to the arising of strong
anisotropy in the pressures, which may produce a transverse
collapse of the star.
\end{abstract}

\section{Introduction}

In the paper \cite{Chaichian} was shown that a relativistic gas
under the influence of a magnetic field $B$ of order of the
quantum electrodynamics  limit of $m_e^2/e \sim 10^{13}$ Gauss and
for densities $N \leq 10^{30}$ cm$^{-3}$ becomes unstable and
collapses, since the pressure perpendicular to the field vanishes.
Physically, the system, infinitely degenerate with regard to the
orbit's center quantum number,  becomes otherwise one dimensional,
with all the electrons falling on the Landau ground state $n=0$. 
The magnetic ``Bohr radius" being of order $\sqrt{\hbar
c/eB}=10^{-2}$\AA. Its spectrum looks equivalent to that of a free
unidimensional particle moving along the external field. under 
these conditions the system is unable to exert any transverse pressure.

Looking at the problem from another side, in a recent paper
\cite{Chakrabarty} has been shown that an electron gas confined
to the Landau ground state cannot be in $\beta$-decay equilibrium
in a neutron star due to an incompatibility among the spin
orientation of the particles involved. The suggestion is given in
\cite{Chakrabarty1} that a bosonization of the electron gas may
take place, which would mean a solution to the above-mentioned
problem. We will discuss here the behavior of the bosonic gas
whose constituent particles are relativistic bosons, which we
suggest may represent the electron pairs, under such extreme conditions
of confinement to the $n=0$ Landau ground state. As is known in
normal superconductors, scalar pairing condensates occur in
absence of a magnetic field (Cooper pairs). The pairing leading to
bosonization is mediated by some interaction as is the Coulomb
exchange interaction among electrons, for instance, in the
ferromagnetic pairing in atoms. In addition to bosonization, 
another effect could be associated with a small change in the new 
particle mass, that is, the final particle would have a mass smaller
than twice that of the initial unpaired particles since the whole
relativistic energy is reduced due to inertial effects. In view of 
this, we shall assume that this binding energy is not significant.

When a magnetic field is applied, the superconductivity
(condensate) is distroyed at some critical magnetic field, (the
Schafroth critical field). But if the magnetic field increases
largely enough  to have a significant fraction of its density in
the Landau ground state, it has been suggested that the condensate
reappears as a consequence of some interaction compatible with a
spin-one vector pairing \cite{Boebinger}. This would lead to a
superconductive-ferromagnetic behavior.

Such particles would  carry twice the charge of the electron and
an effective mass which in principle we take of order twice of
that of the electron, although some corrections must be introduced
however, due to effects coming from the large density and the
magnetic field. Then the system may be treated by following the
same formalism used in previous references \cite{Chaichian,Hugo}. 
As a consequence of condensation the system would
behave as ferromagnetic and under the action of an external field
$H$, a magnetization ${\cal M}$ arises, leading us to define a
microscopic magnetic field $B=H+4\pi {\cal M}$. The interesting
point here is that, due to the positive character  of ${\cal M}$
it may occur that $B \sim 4\pi {\cal M}$, or $H \ll 4\pi {\cal M}$
i.e., the microscopic magnetic field be produced by
self-magnetization.

\section{The condition of self-magnetization}

We shall propose that the relativistic paired electron system
behaves as a vector particle with energy eigenvalues \cite{Hugo,Chaichian} $\epsilon_{0}(p_3) = \sqrt{ p_3^2 c^2 + M^2 c^4 - 2eB \hbar c}$, 
for the Landau ground state $n=0$ (where we take $M=2m_e$, $m_e$ 
being the electron mass), and $\epsilon_{n}(p_3) = \sqrt{
p_3^2 c^2 + M^2 c^4 + 4e B \hbar c (n + \frac{1}{2})}$ for the
excited states $n=1,2...$ We observe that the magnetic field
introduces an effective mass for vector bosons $M_{0}=\sqrt{M^2  -
2e B \hbar /c^3}$ in the ground state such that as $B$ increases,
$M_{0}$ decreases. This leads to an effective magnetic moment in
the ground state ${\it m= e\hbar/2M_0 c}$. The magnetic mass is
$M_{n}=\sqrt{M^2 + 2eB \hbar(n + \frac{1}{2}) /c^3}$ for the
excited states, which increases with $B$ and $n$.

In Ref.\cite{Hugo} we have shown that Bose-Einstein condensation, in
the sense of a large population in the Landau ground state having
its momentum along the magnetic field equal to zero or very small,
occurs for scalar and vector particles in presence of a strong
magnetic field.  We name $n_{0}^{\pm} = [exp(\epsilon_0 \mp
\mu)\beta - 1]^{-1}$ the density of particles and antiparticles,
respectively, in the ground state.
 We expect then most of the population of particles to
be around the ground state, since for small temperatures
$n_{0p}^-$ is vanishing small and $n_{0}^+$ is a bell-shaped curve
with its maximum at $p_3 = 0$. We will define $\mu' = \mu - M_{0}
c^2$ and recall the procedure followed in \cite{Hugo}. We call
$p_0 (\gg \sqrt{- 2 M_{0} \mu'})$ some characteristic momentum.
Taking by symmetry the density of particles minus antiparticles
(the latter will vanish as $-\mu' \ll T$) we have in a small
neighborhood of $p_3 = 0$,

\begin{eqnarray}
N_{0} & = & \frac{2e B T}{2 \pi^2 \hbar^2 c}\int_0^{p_0} \frac{d
p_3}{\sqrt{p_3^2 c^2 + M_{0}^2 c^4 \pm 2e B \hbar c} - \mu}
-\int_0^{p_0} \frac{d p_3}{\sqrt{p_3^2 c^2 + M_{0}^2 c^4 \pm 2e B
\hbar c} + \mu} \nonumber
\\[1em]
&&\mbox{}\simeq \frac{2e B T}{2 \pi^2 \hbar^2 c^2}\int_0^{p_0}
\frac{(M_{0}c^2 + \mu) d p_3}{p_3^2 c^2 +M_{0}^2 c^4 - \mu^2}-
\int_0^{p_0} \frac{(M_{0}c^2 - \mu) d p_3}{p_3^2 c^2 +M_{\pm}^2
c^4 - \mu^2}\nonumber
\\[1em]
&&\mbox{}= \frac{2e B T}{4 \pi \hbar^2
c}\frac{2\mu}{\sqrt{M_{\pm}^2 c^4-\mu^2}} \sim \frac{2e B T}{4 \pi
\hbar^2 c}\sqrt{\frac{2M_{0}}{-\mu'}} \label{5}
\end{eqnarray}

\noindent where $N = N_{0} + \delta N$ and $\delta N$ is the
density in the interval $[p_0, \infty]$. Actually as $\mu'\to 0$,
$N_{0} \to N$ and $\delta N$ is very small. We get then

\begin{equation}
\mu' \simeq - \frac{e^2 B^2 T^2 M_{0}}{2 \pi^2 N^2 \hbar^4 c^2}.
~\label{3}
\end{equation}

We observe that $\mu'$ is a decreasing function of $T$ and
vanishes for $T = 0$, where the "critical" condition $\mu = M_{0}
c^2$ is reached. As shown in \cite{Hugo} in that limit the
Bose-Einstein distribution degenerate in a Dirac $\delta$-function, 
which means to have all the system in the ground state
$p_3 = 0$. From (\ref{5}) we may write the thermodynamic potential
as

\begin{equation}
\Omega= \frac{e B T}{2 \pi \hbar^2 c}\sqrt{M_{0}^2 c^4-\mu^2}
~\label{6}\; ,
\end{equation}

\noindent while the magnetization is given approximately by
\begin{equation} {\cal M} =
-\frac{\partial \Omega}{\partial B} = \frac{e N \hbar}{ M_{0} c}.
\label{mag} \; ,
\end{equation}

\begin{equation}
B = 4\pi{\cal M}= 4 \pi\frac{e N \hbar}{M_{0} c}  \label{consis} \; .
\end{equation}

One can then state the condition for self-magnetization by writing the
equation $H=B-4\pi {\cal M}=0 $. First, let us assume that $N \sim
10^{32}$. Then ${\cal M}\sim 10^{12}$ and $B\sim 10^{13}$~G. Therefore,
the condition for self-magnetization is satisfied. The system becomes a
giant magnet whose stability is determined by the transverse pressure
condition $P_{\perp} \thickapprox P_{3}$, where
$P_{\perp}=-\Omega-B{\cal M}$. However, the estimate value of $\Omega$
would depend on the fraction of electrons paired. Let us name $N_u$,
$N_p=N-N_u$ the density of unpaired and paired electrons, respectively.
If $N_u \sim N_p$. Then the dominating pressure comes from the
(unpaired) electron gas contribution, $\Omega \sim N M_0 \sim 10^{26}$,
which is very close to $B{\cal M} \sim 10^{25}$. Thus, one can affirm
that in the region $N\sim 10^{32}-10^{33}$, the condition
$P_{\perp}<<P_{3}$ holds and the bosonized ferromagnetic white dwarf
star destabilizes and collapses.  If $N_u \ll N_p$ so that te dominant
pressure comes from the paired gas the collapse is also unavoidable
since $P_{\perp}\sim 0$ . As $\Omega$ is positive, its contribution to
pressure is negative. The stability requires from a Fermion
background.

There is still another point to be considered when $eB$ approaches
to $M$. As $M_{0}$ decreases with increasing $B$, the
magnetization ${\cal M}$ increases with $B$, and would diverge for
$M_{0} \to 0$. For $eB\hbar/ c^3$ close enough to $M$ one expects
the main contribution to $B$ be produced by ${\cal M}$.  We get an
equation similar to the one discussed in \cite{Chaichian} for the
$W$ condensate.

Let us write $2eB\hbar/M^2 c^3= x^2  $ where $0 \leq x \leq 1$.
For $x=1$, we have the critical field $B_C =M^2 c^3/2e \hbar
\simeq 8.82 \times 10^{13}$ G. Then we can write
$M_0=M\sqrt{1-B/B_C}$. We easily get
\begin{equation}
x^2\sqrt{1 - x^2} =\frac{8 \pi e^2 \hbar^2 N}{ M^3 c^4} = A.
\label{cub}
\end{equation}

\noindent By simple inspection we find that it has real solutions
only for $A<2\sqrt{3}/9=A_1$. This means that $N \leq 10^{32}$.
By solving the cubic equation (\ref{cub}) we find that for $A \ll
1$ these real solutions are $x_1 = \sqrt{A + A^2/2}$ and $x_2 =
\sqrt{1 - A^2}$. The first solution means that $B$ increases with
increasing $N$, up to the value $B_{max} = 2 /3B_C$. In the
second solution $B$ decreases for growing  $N$, and its limit for
$N \to 0$ being $B_C$. The last result has only meaning if
interpreted as indicating that the expression for the
magnetization (\ref{mag}) is incomplete. Actually, it must include
the contribution from Landau states other than the ground state,
which lead to a diamagnetic response to the field. The decreasing in
population of the ground state is compensated by the increasing of 
the number of particles in Landau states with $n>0$. Their
contribution would compensate the increasing of the self-consistent
field with increasing $N$ to keep $B < B_C$.

\section{Conclusions}

From all the previous reasoning we conclude that an  electron
system, as a white dwarf, can be hardly stable at fields of order
$B_c$. In principle, such fields can be maintained
self-consistently, but the possibility of a collapse is highly
increased: the one-dimensional world created by the magnetic field
is completely unstable. The previous results, if applied to the
condensation of $\rho$ and $\omega$ mesons (for instance, in
neutron stars), would be compatible with the self-magnetization
condition for densities of order $10^{39}$~cm$^{-3}$ if the
resulting magnetic fields are in the interval $10^{17}-10^{19}$~G.
Instabilities would lead to a collapse. This point deserves
further research, as it provides additional arguments to those
given in \cite{aurora} against the claimed existence of the
so-called magnetars, because those objects seem to be unstable 
under such $\rho$ and $B$ conditions.

\end{document}